\documentclass[namedreferences]{solarphysics}
\usepackage[optionalrh,solaromanenum]{spr-sola-addons} 
\usepackage{graphicx}                    
\usepackage{color}                       
\usepackage{url}                         
\usepackage{hyperref}
\ifx \doiurl    \undefined \def \doiurl#1{\href{http://dx.doi.org/#1}{\textsf{DOI}}}\fi
\ifx \adsurl    \undefined \def \adsurl#1{\href{http://adsabs.harvard.edu/abs/#1}{\textsf{ADS}}}\fi
\ifx \arxivurl  \undefined \def \arxivurl#1{\href{http://arxiv.org/abs/#1}{\textsf{arXiv}}}\fi

\newcommand{\etal}{{\it et al.}}

\newcommand{\aap}{    {\it Astron. Astrophys.}}

\newcommand{\apj}{    {\it Astrophys. J.}}
\newcommand{\apjl}{   {\it Astrophys. J. Lett.}}

\newcommand{\solphys}{{\it Solar Phys.}}

\begin{document}

\begin{article}

\begin{opening}

\title{On the Long-Term Modulation of Solar Differential Rotation}

%
\author{M.~\surname{Suzuki}$^{1}$    
       }

%
\runningauthor{M. Suzuki}
\runningtitle{Long-Term Modulation of Differential Rotation}

%
  \institute{$^{1}$ Mie University, Tsu-City, Mie Prefecture, 514-8507, Japan
                     email: \url{suzuki-obs@pop02.odn.ne.jp}
             }

\begin{abstract}
Long-term modulation of solar differential rotation was studied with data from Mt. Wilson and our original observations during Solar Cycles 16 through 23. The results are:
\romannumeral1 ) The global {\it B}-value ( {\it i.e.} latitudinal gradient of differential rotation), is modulated in a period of about six or seven solar cycles. 
\romannumeral2 ) The {\it B}-values of the northern and southern hemispheres are also modulated with a similar period to the global one, but 
\romannumeral3 ) they show quasi-oscillatory behavior with a phase shift between them. 
  We examined the yearly fluctuations of the {\it B}-values in every solar cycle with reference to the phase of the sunspot cycle and found that the {\it B}-values show high values over the full-cycle years, when the cycle-averaged  {\it B}-values are high. 
  We discuss the independent long-term behavior of solar differential rotation between the northern and southern solar hemispheres and its implication for the solar dynamo.
\end{abstract}

%

\end{opening}

%
 \section{Introduction}
 
 As the solar differential rotation is believed to play an important role in driving cyclic solar activity, it is important to study closely the characteristics of differential rotation of the Sun, which may give us some clues for understanding the origin of solar activity. However, we have not yet obtained conclusive results on the characteristics of differential rotation. What is the possible period of long-term modulation? How does the differential rotation depend on the sunspot activity? What is the nature of the North--South asymmetry of differential rotation? Although all of these issues are essential for our understanding of the maintenance of differential rotation and dynamo theory, they have not yet been solved, probably due to the difficulties of extracting signals from fluctuating motions of rotation tracers or from Doppler velocity data in the presence of complex convective or oscillatory motions in the solar atmosphere. \\
 
Longitudinal motions of sunspots are the only available data for the investigation of long-term variation of differential rotation. Its latitudinal dependence is usually expressed as $\omega$\, = {\it A} - {\it B}\,$\sin^{2}\phi$ (Newton--Nunn relation), 
where $\omega$ is the rotation rate and $\phi$ is the solar latitude.  \inlinecite{JBU05} showed that the {\it B}-value averaged over one solar cycle ( {\it i.e.} the cycle-averaged {\it B} ) varies from cycle to cycle. \inlinecite{JBU05} and \inlinecite{Suzuki12} showed that the cycle-averaged {\it B} varies quasi-periodically over a long-term span of eight or nine cycles, similar to the Gleissberg cycle found for sunspot number. \inlinecite{Zhang13} found a similar quasi-periodic variation of 80\,--\,90 years in the rotation rate of a latitude zone where sunspots are found to be prolific. At present, the value of the period seems to depend on the averaging time span and is open for further study.\\

Several authors have reported the dependence of differential rotation on sunspot activity.  \inlinecite{Bal80} showed that the solar rotation tends to be less differential to the end of an activity cycle with the Greenwich sunspot data (1940\,--\,1968).  \inlinecite{NRibes93}  similarly showed that the rotation is more differential at the solar minimum and is less differential at the solar maximum from Meudon data (1977\,--\,1984).   With much  longer Greenwich data (1874\,--\,1976), \inlinecite{JG95} confirmed the results of previous authors. Further more, \inlinecite{KHH93} obtained the same result with their 17 years magnetogram data. 
On the contrary, \inlinecite{Lustig83} argued that the solar rotation is less differential at the activity minimum and more differential at the activity maximum from their sunspot observations at Kanzelh\"{o}he. On the other hand, we have more than three cycles of the Mt. Wilson Doppler and helioseismic observations of the `torsional oscillation', which is a sunspot-cycle modulation of the differential rotation with a period of 11 years, drifting from the poles to the equator in 22 years (\opencite{Howard80}; \opencite{Howe09}).
So we do not yet have an observationally conclusive result on the  relation between sunspot activity and the differential rotation. \\

Solar active phenomena and relative sunspot numbers have been known to show asymmetry between the North and South hemispheres ({\it cf.} \inlinecite{JG97} and references cited therein). \inlinecite{JG97} analyzed the Greenwich sunspot data (1879\,--\,1976) and found that the equatorial  rate ({\it A}-value) and the  latitudinal gradient ({\it B}-value) of rotation  have several peaks of oscillatory power, ranging from 45.5 through 10.5 years.  They further suggested the possibility of an antisymmetric torsional oscillation of the Sun. Recently, \inlinecite{Zhang13}, with the analysis of the Greenwich data and USAF/NOAA (1877\,--\,2008), found evidence of anti-correlation of the rotation in the two hemispheres and oscillatory behavior of the asymmetry  at a period of 80\,--\,90 years. However, previous results do not seem to be consistent, probably due to the difficulty of determining the rotation parameters in sunspot minimum phases, where numbers of sunspots may be very low. Careful treatment seems necessary to obtain conclusive results on the North--South asymmetry of the solar rotation.\\

As mentioned above,  \inlinecite{Suzuki12} reported that the solar {\it B}-value increased from Cycle 22 to 23 with his original observation of sunspots.  We more closely analyzed Suzuki's original data, searched for the reason for the cyclic {\it B}-value modulation, and found that the temporal variation of the North--South asymmetry of the {\it B}-value in sunspot minimum phases may be the source of the {\it B}-value cyclic modulation. By extending our analysis to data with Mt. Wilson and Greenwich sunspot data, we confirmed that our finding is valid for data with a much longer time span.  Moreover, we found that the {\it B}-values not only in the sunspot minimum but also in the sunspot maximum phases of the solar cycles contribute, with nearly the same weights, to the cycle-average {\it B}-value. In other words, the {\it B}-values both in the activity minimum and in the maximum phases in a cycle are high when the cycle-averaged {\it B}-value is high. Our analysis on the North--South asymmetry of solar rotation suggests that the {\it B}-values in the two hemispheres show oscillatory behavior with nearly the same period of Gleissberg cycles, but with their phases shifted relative to one another. 

\section{Differential Rotation in Cycles 22 and 23}

The sunspot data used in this section are the same as that which was reported by \inlinecite{Suzuki12}, where his observation, measurements,  and reduction procedures for sunspot rotation are described in detail.\\

  \begin{figure}    
   \centerline{\includegraphics[width=1.0\textwidth,clip=]{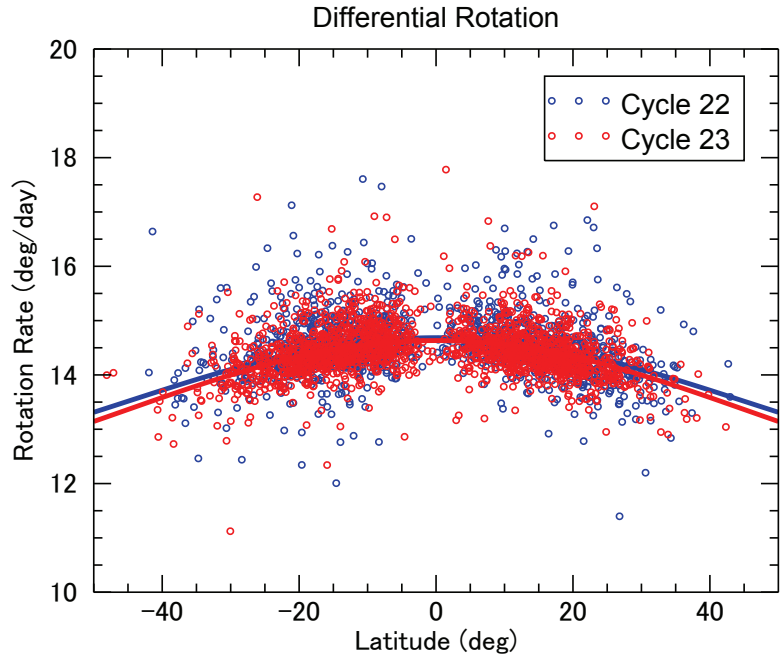}
              }
              \caption{Latitudinal dependence of sunspot rotation in Cycle 22 and Cycle 23 with Newton--Nunn least-square fitting curves.
                      }
   \label{Figure1}
   \end{figure}
   
\subsection{Cycle-to-Cycle Variation}

The cycle-averaged {\it B}-values showed a monotonically decreasing trend from Cycles 18 through 22, while that of Cycle 23 clearly increased significantly (\opencite{JBU05}; \opencite{Suzuki12}).  
In both works, the {\it B}-values are derived by least-square fitting with the Newton-Nunn function to the 5$^{\circ}$ binned sunspot rotation data in latitude.\\ 

To see the possible influence of binning on the previous conclusion, we performed the same reduction on the same data without 5$^{\circ}$ binning.
In Figure 1, we plot the rotation rates along the latitudinal locations of all the sunspot groups observed and measured by us during Cycles 22 and 23.  Newton--Nunn fitting curves are over-plotted at the two cycles, separately. The fitting curve for Cycle 23 is more convex to the top than that of Cycle 22, which means a higher {\it B}-value in Cycle 23. Actually the {\it B}-value is $2.340{\pm}0.077$ for Cycle 22 and is $2.537{\pm}0.051$ for Cycle 23.  We obtained essentially the same result as before by not taking latitudinal data binning, and so confirmed our previous conclusion. \\

   \begin{figure}    
   \centerline{\includegraphics[width=1.0\textwidth,clip=]{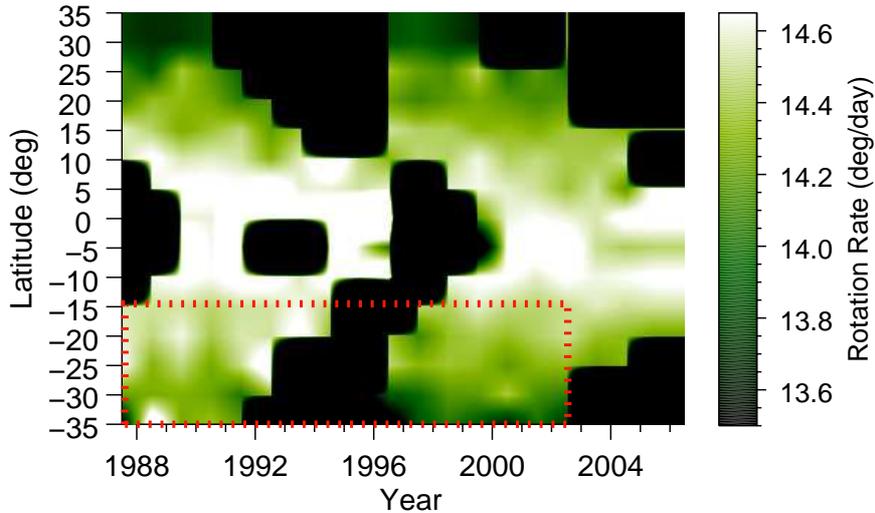}
              }
              \caption{Yearly variation of sunspot rotation in Cycle 22 and Cycle 23 for 5$^{\circ}$ latitudinal zones  (Butterfly-diagram format). 
                      }
   \label{Figure2}
   \end{figure}

\subsection{Yearly Variation of Rotation}

We closely studied the yearly variation of solar rotation to see at which phase in a cycle the {\it B}-value shows high or low values, which may lead to the difference of the cycle-averaged {\it B}-values.  In Figure 2, we show the yearly variation of rotational speed during the period of Cycles 22 and 23 for each latitudinal zone of 5$^{\circ}$ width. The figure is a butterfly-diagram of rotational speed.  In the figure, rotation rates at high latitudes in the northern hemisphere do not show any significant difference between Cycles 22 and 23.  
In low-latitude zones, there may be a slight difference of rotation speed between the two cycles. On the other hand, we notice a significant difference in the southern hemisphere, where rotation speeds in higher-latitude zones such as 20\,--\,25, 25\,--\,30 
and 30\,--\,35$^{\circ}$ show lower rotation rates in Cycle 23 than in Cycle 22 as is shown with a red box in the figure. We may consider that the difference of rotation in the southern hemisphere may be the source of the high {\it B}-value in Cycle 23. \\

As high latitude sunspots generally emerge in early phases of solar cycles, we may expect that the difference in the cycle-averaged {\it B}-value is due to the difference of rotation at initial phase of every cycle. \\

In this section, we obtained a few hints of the North--South asymmetry and the cycle-to-cycle variation of differential rotation by the comparison between Cycles 22 and 23. Further analysis on these topics will be done in the next section with a much longer sunspot dataset. \\
%
\section{Further Analysis in a Much Longer Time Span}

We studied the North--South asymmetry of solar rotation {\it B}-value for a much longer time span with the published sunspot rotation data by \inlinecite{Howard84} for 1921\,--\,1982, Mt.Wilson sunspot group data supplied at the NOAA site (\href{ftp://ftp.ngdc.noaa.gov/SOLAR_DATA/SUNSPOT_REGIONS/Mt_Wilson/}
{ftp:\slash\slash{}ftp.ngdc.noaa.gov\slash{}SOLAR\_DATA\slash{}SUNSPOT\_REGIONS\slash{}Mt\_Wilson\slash{}}) 
for 1983\,--\,1987, and Suzuki's observation for the years 1988\,--\,2012. 
Both the Howard data and Mt. Wilson sunspot group data are based on the Mt. Wilson direct photographs of the Sun.
\inlinecite{Howard84} gave the yearly averaged rotation rate for every 5$^{\circ}$ latitudinal bin. Therefore, we performed the same averaging and compiling process for the Cycles 21\,--\,23 data as \inlinecite{Howard84}. We will base our analysis on this extended time span in the following. Latitudinal binning, we think, has little influence on the derived results, as is confirmed in Section 2.1.  
 
\subsection{Cycle-to-Cycle Variation of North--South Asymmetry of {\it B}-Value}

At first, we studied the cycle-to-cycle variation of {\it B}-values. The {\it B}-values are derived for the northern and southern hemispheres, separately, assuming the Newton--Nunn relation. We also derived the global {\it B}-values with the rotation-rate data of both hemispheres. Our method to derive the {\it B}-value for each solar cycle is as follows: \romannumeral1) One solar cycle includes around 11 years' data. We derived average rotation rate for every latitude bin, by taking the arithmetic average of yearly rotation rates of the years in the cycle. \romannumeral2) The cycle-averaged rotation rates thus derived were fitted to the Newton--Nunn form to estimate the {\it B}-value and its standard deviation.\\

   \begin{figure}    
   \centerline{\includegraphics[width=1.0\textwidth,clip=]{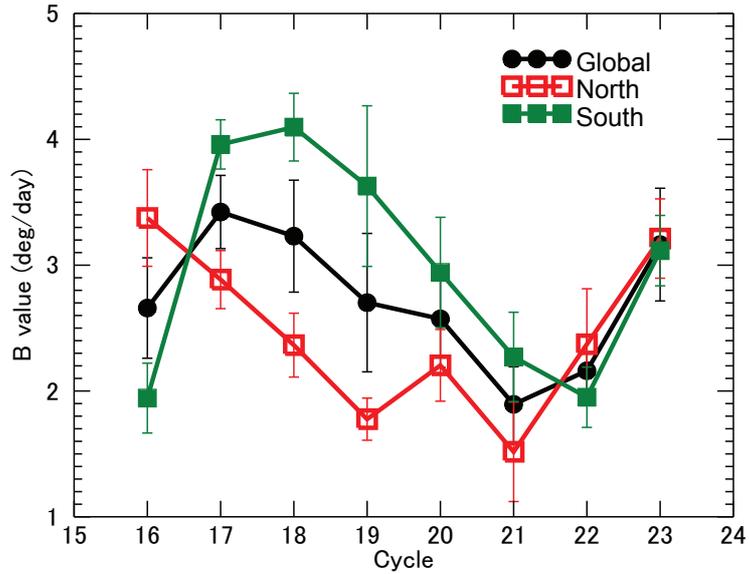}
              }
              \caption{Variation of the cycle-averaged {\it B}-values for the northern (orange) and southern hemispheres (green) separately, and for the global sphere (black).  
                      }
   \label{Figure3}
   \end{figure}

Figure 3 shows that the {\it B}-values of the northern and southern hemispheres and those of the global sphere, respectively, vary in a  quasi-oscillatory fashion over a long time span. Each {\it B}-value oscillates with a period of around six or seven solar cycles, although the oscillation phases are shifted to one another. The phase difference between the North and the South leads to the change of the North--South asymmetry change: The northern {\it B}-value is higher during Cycle 16, becomes lower than that of the South during  Cycles 17\,--\,21, and then resumes higher values for Cycles 22\,--23\,.\\

The global {\it B}-value comes from the averaging of the both hemispheres' {\it B}-values. So on the long-term modulation 
of the global {\it B}-value studied by \inlinecite{JBU05} and \inlinecite{Suzuki12}, we may conclude that the global {\it B}-value variation is due to the phase-shifted quasi-oscillation of the {\it B}-values in the northern and southern hemispheres.\\

  It is worthwhile here to examine the possible errors in our results. Our sunspot-group data for Cycles 16\,--\,20 are those of Mt. Wilson data available on-line. By comparison between the Mt. Wilson data and Kodaikanal data, \inlinecite{Howard99} suggested that systematic errors in rotation-rate measurements may give rise to false North--South asymmetry of solar rotation. They found that the sunspot-rotation data of both sites have systematic errors due to small mis-identification of solar-rotation axes of $\pm{0.4}^{\circ}$. They also found that subtle optical effects at both observatories introduce additional systematic errors 
to the sunspot-rotation data. According to their best error estimation and reduction of the Mt. Wilson data, the {\it B}-value for the years 1917\,--\,1985 was corrected from $2.99\pm{0.06}$ to $3.02\pm{0.06}$ (deg\,day$^{-1}$), {\it i.e.}, a correction of 1{\,\%}. The variation in the North--South asymmetry obtained in the present article is well above the 1{\,\%} correction.  Suzuki's observation and reduction were carefully done as to the identification of solar-rotation axis as was described by \inlinecite{Suzuki98} and \inlinecite{Suzuki12}. He obtained the rotation-rate by tracking those sunspots located within a central meridian distance of 60$^{\circ}$ to avoid the possible influence due to the Wilson depression effect and subtle optical astigmatism. We think that the smooth variation of North--South asymmetry shown in Figure 3 indicates a similar quality of reduction for the Mt. Wilson and Suzuki data.\\

\subsection{Close Examination of the Cyclic Modulation of the {\it B}-Value}

To display the temporal variation of the {\it B}-value in more detail, we studied the behavior of the yearly {\it B}-values during the period 
from 1923 through 2012 ({\it i.e.} Cycles 16 to 23).  Figure 4 shows the variation of the {\it B}-values of the global sphere and of the northern and southern hemispheres, respectively. The increase of the {\it B} in the early and final phases of sunspot cycle is sharp, as has already been pointed out by many authors (\opencite{Bal80}; \opencite{NRibes93}; \opencite{KHH93}; 
\opencite{JG95}; \opencite{Suzuki12}).  On the other hand, the {\it B}-values in the mid-term years of a sunspot cycle (hereafter referred as {\it $B_M$}) do not show large fluctuations in the cycle but change gradually from cycle to cycle. \\

The {\it $B_M$} values of each cycle are derived with the same method as is described in the previous section, but with the data only for midterm years.  Midterm years for each cycle are indicated in Figure 4. The variations of {\it $B_M$} are plotted in Figure 5.  
The {\it $B_M$} shows similar variation as the cycle-averaged {\it B}, which means that the variation of {\it $B_M$} is one of the sources of cyclic variation of the cycle-averaged {\it B}-values.  The phase shift between the north and the south {\it $B_M$}s is also very akin to the cycle-averaged {\it B} oscillation.\\

However, there are some differences in the variation characteristics between the {\it $B_M$} and the cycle-averaged {\it B}. The {\it $B_M$} values in the southern hemisphere are lower in Cycles 19\,--\,20 than the cycle-averaged {\it B}. The increase of {\it $B_M$} from Cycle 22 to 23 is not so steep as the cycle-averaged {\it B}. To search for another source of the cycle-averaged {\it B} variation, we examined the correlation between the cycle-averaged {\it B} with the maximum {\it B} in a cycle ({\it i.e.} the ``sharply increased component'' mentioned above) and with the minimum {\it B}, separately in the global sphere, and the northern and southern hemispheres, as is shown in Figure 6. The correlations between the cycle-averaged {\it B} and the {extreme} values of {\it B} in a cycle are positively correlated, although the minimum {\it B} values in the southern hemisphere have less systematic relation to the cycle-averaged {\it B}.  So we may consider that the extreme {\it B} values often seen in the start or end years of solar cycles are also the source of cyclic variation of the cycle-averaged {\it B}-values.\\

Therefore, we consider that not only the {\it B}-values in the midterm phase but also those in the start and end phases contribute to the cycle-averaged {\it B}-value, irrespective of the global, northern or southern hemispheric ones.\\
 
   \begin{figure}    
   \centerline{\includegraphics[width=1.0\textwidth,clip=]{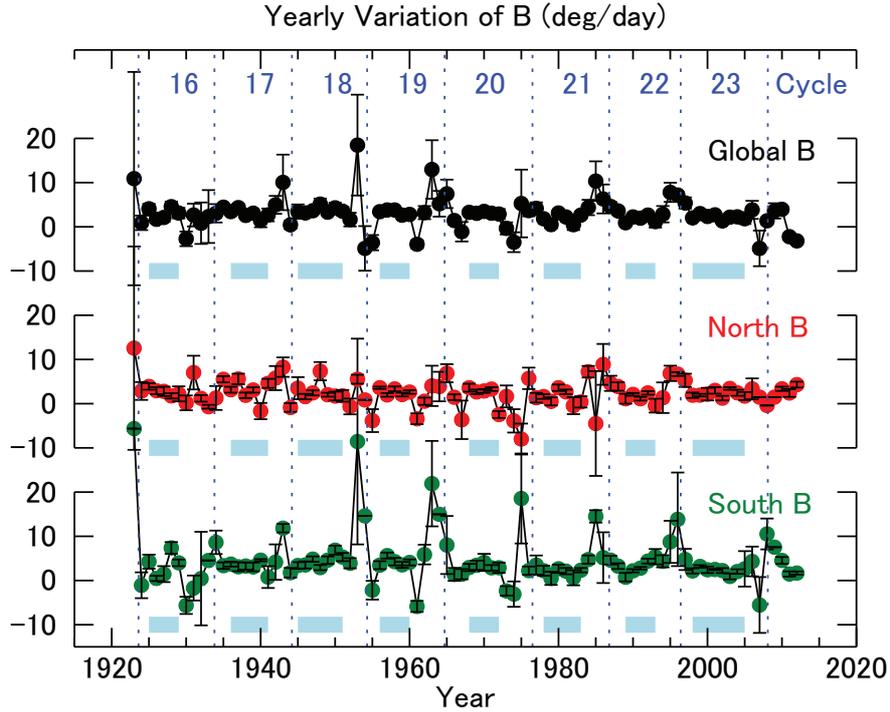}
              }
              \caption{Variation of the yearly {\it B}-values for northern (orange) and southern hemispheres (green) separately, and for the global sphere (black). Midterm years of solar cycles are indicated by horizontal light-blue bars. 
                      }
   \label{Figure4}
   \end{figure}
   \begin{figure}    
   \centerline{\includegraphics[width=1.0\textwidth,clip=]{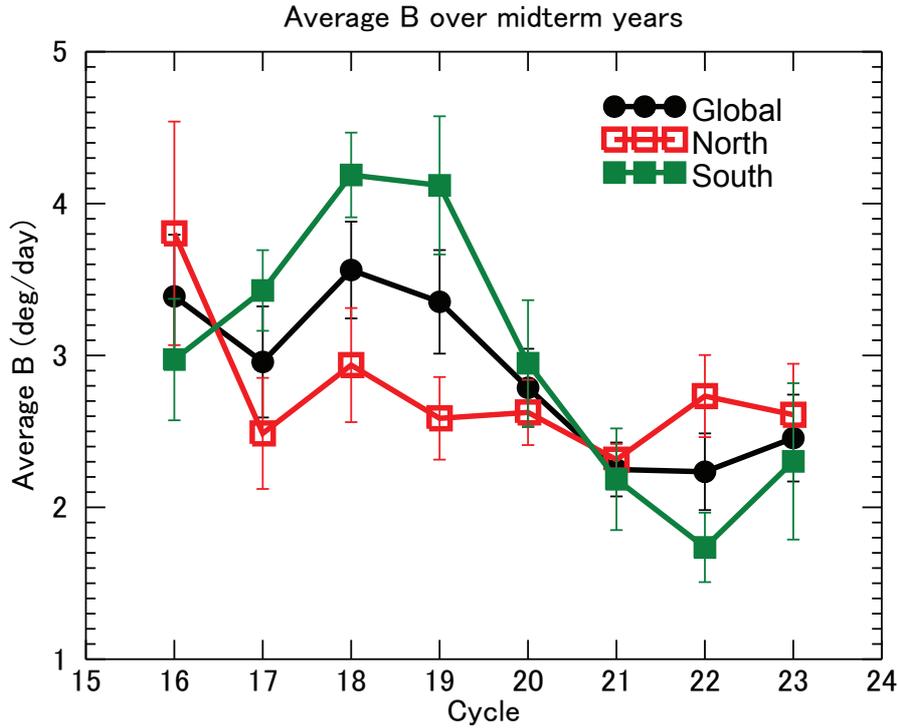}
              }
              \caption{Variation of the midterm-averaged {\it B}-values for northern (orange) and southern hemispheres (green) separately, and for the global sphere (black).  
                      }
   \label{Figure5}
   \end{figure} 
  
      \begin{figure}    
   \centerline{\includegraphics[width=1.0\textwidth,clip=]{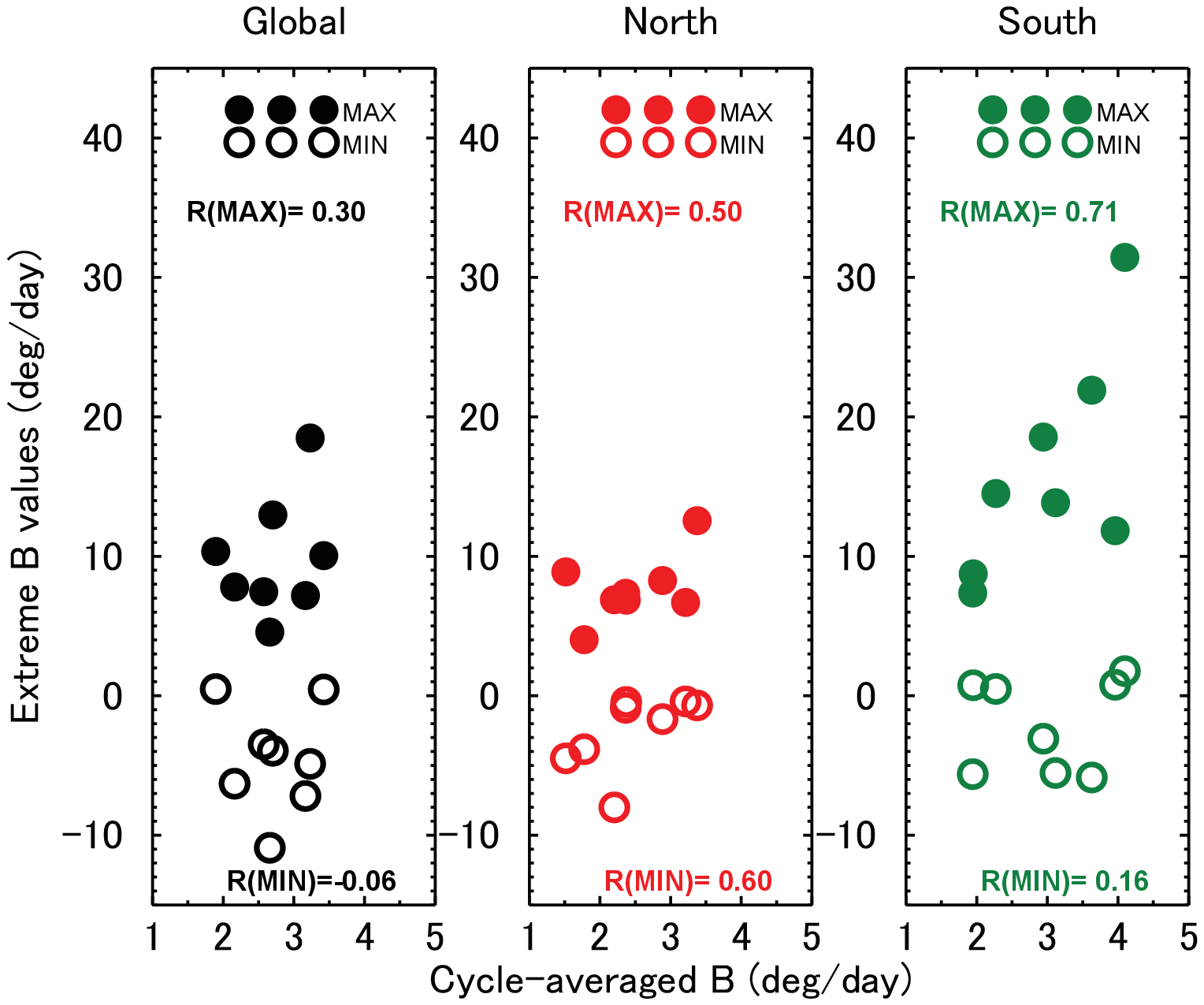}
              }
              \caption{Correlation between the cycle-averaged {\it B}-values and the extreme (maximum or minimum) {\it B}- values of every cycle.  The maximum or minimum {\it B}-values occur at the beginning or end years of cycles ({\it i.e.} when sunspot numbers are low). {\it R(MAX)} is the correlation coefficient between the maxima and the cycle-averaged {\it B}-values. {\it R(MIN)} is that for the minima.
                      }
   \label{Figure6}
   \end{figure}
   
\section{Conclusion and Discussion}

To summarize this work : \romannumeral1) The global {\it B}-value is modulated with a period of about six or seven solar cycles. \romannumeral2) The {\it B}-values of the northern and southern hemispheres are also modulated with a similar period to the global one. \romannumeral3) The long-term variation of the {\it B}-value in each hemisphere shows quasi-oscillatory behavior with a phase shift relative to one another. 
\romannumeral4) The yearly variation of the {\it B}-value in one solar cycle has different behavior between in the midterm years ({\it i.e.} sunspot-maximum years) and in the start and end years of one cycle  ({\it i.e.} sunspot-minimum years ). In the latter case, the {\it B}-values have large and erratic variation, while those in the former case show little fluctuations. 
\romannumeral5) While the yearly variation of {\it B}-value fluctuates differently depending on the phase in a sunspot cycle, the {\it B}-values over all the years in a sunspot cycle contribute to the cycle averaged {\it B}-value variation, irrespective of the north or south hemispheric ones.\\ 

\inlinecite{Zhang13} derived long-term modulation of North--South asymmetry of solar rotation with a period of about 80\,--\,90 years. Our result on the long term modulation is consistent with their result, although the quantitative estimate of the period is different to one another. The difference is probably due to the temporal resolution of rotation analysis. In any case, the oscillatory behavior of the North--South asymmetry will prevail on the Sun. Our new proposal is that the northern and southern hemispheric differential rotations show quasi-oscillatory variation with a phase shift relative to one another. The long-term variation of the North--South asymmetry of the differential rotation found in \inlinecite{Zhang13} and confirmed in this work, is due to the phase-shifted oscillation of both hemispheres. \\

According to recent works on the solar dynamo and rotation (\opencite{Miesh08}; \opencite{Hotta10a}, \citeyear{Hotta10b}), the solar differential rotation can be considered as maintained by the supply of the angular momentum from the inner core through the tachocline, and by the transport and distribution of the angular momentum over the solar convective zone by meridional circulation and/or global convection.  Our results imply that the solar northern and southern hemispheres may receive and transport the angular momentum differently from each other, while the basic maintenance mechanism is common. We may speculate that the North--South asymmetry of the solar features, such as relative sunspot number {\it etc.}, may be due to the asymmetry of the differential rotation through the $\Omega$-effect in dynamo theory. \\

Recently \inlinecite{McIntosh13} suggested that the the Sun's hemispheres are significantly out of phase with one another. Their study of historical sunspot records shows that the hemispheric ``dominance'' has changed twice in the past 130 years. Our results support their physical picture on the solar cyclic variation, although epochs of ``dominance'' change (in the middle of Cycle 16 and around the minimum between Cycles 19 and 20) does not always coincident to those of the differential rotation change. The question remains open for further non-linear theoretical study of solar dynamo.\\
 
Concerning the general trend that the sunspot rotation is more differential in the sunspot minimum than in the sunspot maximum, a possible physical origin of the trend was speculated on by previous researchers. The difference of anchoring depth of young active regions during a cycle was proposed by \inlinecite{Bal80}, the reduction of positive internal shear of deep convective layers by 
\inlinecite{NRibes93}, the difference of anchoring depth of rotation tracers in slowly varying convective flow field by \inlinecite{KHH93}, and the planetary configurations by \inlinecite{JG95}. We think that it remains a open question at present without a conclusive agreement.  Numerical-simulation studies and helioseismological observational studies of convection zone are needed. \\

In addition to the previous issue, the direct confirmation of North--South asymmetry of differential rotation is strongly needed. Helioseismology discloses the internal structure and rotation field in the convective zone of the Sun (\opencite{Schou98}). However, their analyses and succeeding researches have assumed North--South symmetry of the rotation field. Large-scale time-distance 
analysis of the solar deep layers has mostly focused on the North--South symmetrical meridional flow so far. Since evidence of North--South asymmetry now has been found in several parameters of the Sun, it seems very interesting and valuable to do helioseismological analysis of the actual Sun's interior without the assumption of North--South symmetry, which will give us more clues for the understanding of the solar dynamo.\\

%

%
\begin{acks}
The author acknowledges J. Kubota and R. Kitai for their fruitful discussions and helpful advices to the research. He also thanks M. Hagino for giving valuable information on the North--South asymmetry of sunspot magnetic helicities. Finally he thanks the  anonymous referee for suggesting important improvements to clarify the presentation of the article.
\end{acks}

%
%
%

\end{article} 
\end{document}